\documentclass[a4paper,10pt]{article}
\usepackage{pos}
\usepackage[utf8]{inputenc}
\usepackage{lineno}
\usepackage[utf8]{inputenc}
\usepackage{xcolor}
\usepackage{graphicx}
\usepackage{natbib}
\usepackage{float}
\usepackage{tikz}
\usepackage{amsmath}
\usepackage{import}
\usepackage{setspace}

\begin{document}
\title{Variations to the z-Expansion of the Form Factor Describing the Decay of B Mesons}
\ShortTitle{Variations of the z-Expansion in B Meson Decays}
\author*[a]{Daniel Simons}
\author[b]{Erik Gustafson}
\author[a]{Yannick Meurice}
\affiliation[a]{Department of Physics and Astronomy, The University of Iowa\\
Iowa City, IA 52242, USA}
\affiliation[b]{Fermilab, Batavia, IL 60510, USA}
\emailAdd{daniel-simons@uiowa.edu}
\emailAdd{egustafs@fnal.gov}
\emailAdd{yannick-meurice@uiowa.edu}

\abstract{We examine the decay rate of the particle decay $B^0 \rightarrow D^- \ell^+ \nu_{\ell}$ using data collected from the Belle Collaboration~\cite{mainb2d}. We studied three parameterizations of the form factor which describe the differential decay rate, the Caprini, Lellouch, and Neubert (CLN) parametrization, the Boyd, Grinstein, and Lebed (BGL) parametrization, and the Bourrely, Caprini, and Lellouch (BCL) parameterization. The form factor is a function of the hadronic recoil variable $w$, and each parameterization contains unique free parameters which are the focus of this work. We test the extrapolations of the form factor by fitting many different subsets of the low $w$ data and then compare the prediction of the fit to the high $w$ data using a $\chi^2$-metric. By only fitting the low $w$ data we are able to examine the stability of extrapolations which will be informative for lattice simulations.}


\FullConference{%
 The 38th International Symposium on Lattice Field Theory, LATTICE2021
  26th-30th July, 2021
  Zoom/Gather@Massachusetts Institute of Technology
}

\maketitle

\section{Introduction}
\label{sec:Introduction}
\noindent

The different parameterizations that were considered were the Boyd, Grinstein, and Lebed (BGL) parameterization~\cite{originalBGL}, the Bourrely, Caprini, and Lellouch (BCL) parameterization~\cite{originalBCL}, and the Caprini, Lellouch, and Neubert (CLN) parameterization~\cite{originalCLN}. The BGL and BCL parameterizations both incorporate the z-expansion, while the CLN does not. The form factors, and ultimately the differential decay rate, are functions of the lepton momentum transfer variable $q^2$. A problem with this is that there is a branch cut in the complex $q^2$ plane. To get form factors that are analytic in the first sheet, a new variable is needed. The z-expansion maps the entire $q^2$ plane onto the unit disk in the complex-$z$ plane, with the branch cut being mapped onto the boundary of the disk~\cite{zexpansiondef}. This allows the form factors to be written as a simple power series of $z$.

These different parameterizations of the form factor are compared by fitting the free parameters to differential decay data obtained by the Belle Collaboration~\cite{mainb2d}. These fits are performed using a limited number of data points in order to test how well these parameterizations are able to predict the remaining data points. The accuracy of the fits and the predictions are then checked by finding the reduced-$\chi^2$ values. Then we compare the parameterizations for both fitted and predicted regions. 

The data that we used had the differential decay rate as a function of the hadronic recoil variable $w$, a function of $q^2$ defined in equation (1). The lowest few data points in $w$ were of particular interest because Lattice QCD calculations are typically restricted to the large $q^2$, which corresponds to the low $w$ values~\cite{Bigi_2016}. This is because Lattice QCD calculations rely on Monte Carlo simulations which cannot reliably extract the values of the form factors at low $q^2$. So the effort to fit only a limited number of data points in my fits is to replicate the conditions of lattice gauge theorists in order to determine the best form factor for them to use.

\section{Theory}
\label{sec:Theory}
\noindent
The kinematics of the decay, $B \rightarrow D l \nu_l$, can be described by the recoil variable $w$. $w$ is defined with the 4-momenta of the $B$ and $D$ mesons $P_B$ and $P_D$ respectively, and $q^2$~\cite{mainb2d}:

\begin{equation}
    w = \frac{P_B \cdot P_D}{m_B m_D} = \frac{m_B^2 + m_D^2 - q^2}{2 m_B m_D}
\end{equation}
\noindent
The quantities $m_B$ and $m_D$ are the rest masses of the $B$ and $D$ mesons respectively. The minimum recoil value corresponds to the $D$ meson having no momentum in the rest frame of the $B$ meson, $w = 1$. The maximal value of $w$ corresponds to when there is no 4-momentum transfer to the lepton-neutrino system, $q^2 = 0$, which leads to $w_{max} = \frac{m_B^2 + m_D^2}{2 m_B m_D} \approx 1.6$~\cite{mainb2d}.

According to the Heavy Quark Effective Theory (HQET), the decay rate of $B \rightarrow D l \nu_l$ can be described, up to a small electroweak correction by~\cite{HQET}:

\begin{equation}
    d \Gamma \propto G_F^2 |V_{cb}|^2 |L_\mu \langle D | \bar{c} \gamma^\mu b | B \rangle|^2
\end{equation}
\noindent
Where $G_F$ is the Fermi coupling constant, $V_{cb}$ is an element of the Cabibbo–Kobayashi–Maskawa matrix, and $L_{\mu}$ is the leptonic current. The hadronic current can be broken down in terms of vector and scalar form factors $f_+(q^2)$ and $f_0(q^2)$ respectively~\cite{mainb2d},

\begin{equation}
    \begin{split}
      \langle D | \bar{c} \gamma^\mu b | B \rangle =
      & f_+(q^2) [(P_B + P_D)^\mu - \frac{m_B^2 - m_D^2}{q^2} q^\mu ] + f_0(q^2) \frac{m_B^2 - m_D^2}{q^2} q^\mu
    \end{split}
\end{equation}
\noindent
In the limit of negligible lepton masses, the differential decay rate does not depend on $f_0(q^2)$ and can be written,

\begin{equation}
    \begin{split}
        \frac{d \Gamma}{dw} =& \frac{G^2_F m^3_D}{48 \pi^3} (m_B + m_D)^2  (w^2 - 1)^{3/2} \eta^2_{EW} |V_{cb}|^2 |G(w)|^2
    \end{split}
\end{equation}
\noindent
This new form factor $G(w)$ has several different parameterizations available that we considered.

A useful, model-independent parameterization of this form factor which only relies on QCD dispersion relations is called the BGL parameterization. With this parameterization~\cite{originalBGL}, 

\begin{equation} 
    \begin{split}
        G(z)^2 & = \frac{4r}{(1+r)^2} f_+(z)^2 \\
        f_+(z) & = \frac{1}{P_+(z) \phi_+(z)} \sum^N_{n=o} a_{+,n} z^n \\
        \phi_+(z) & = 1.1213 (1+z)^2 (1-z)^{1/2}  [(1+r)(1-z) + 2 \sqrt{r} (1+z)]^{-5}.
    \end{split}    
\end{equation}
\noindent
The coefficients $a_{+,n}$ must satisfy the unitarity bound, 
\begin{equation}
    \label{eq:unitarityBGL}
    \sum_{n = 0}^{\infty} |a_n|^2 \leq 1.
\end{equation}
This bound is the result of constraints imposed by QCD.
The new conformal mapping variable, $z$, maps $w$ on to the unit disk in the complex plane \cite{PhysRevD.3.2807, PhysRevD.4.725}. In terms of $w$,  
\begin{equation}
    z(w) = \frac{\sqrt{w+1} - \sqrt{2}} {\sqrt{w+1} + \sqrt{2}}.
\end{equation}

$\phi_+(z)$ is an outer function, which is arbitrary but must be analytic. Following~\cite{mainb2d}, we take the Blaschke factor, $P_+(z) = 1$. The free parameters of this parameterization are the $a_{+,n}$ terms, and there are $N+1$ of them where $N$ is the maximal order of the series.

Another model-independent parameterization of this form factor is called the BCL parameterization. This parameterization has the same form of the form factor $G(z)$, but the vector form factor $f_+(z)$ is now,~\cite{originalBCL}:

\begin{equation} 
    \begin{split}
        f_+(z) & = \frac{1}{P_+(z) \phi_+(z)}  \sum^{N-1}_{n=o} b_{+,n} [z^n - (-1)^{n-N} \frac{n}{N} z^{N}]. 
    \end{split}    
\end{equation}
\noindent
Similar to the BGL parameterization, the coefficients $b_n$ must satisfy a different constraint
\begin{equation}
    \label{eq:unitarityBCL}
    \sum_{j=0}^{N}\sum_{k=0}^{N}b_j b_k B_{jk} \leq 1.
\end{equation}
The symmetric matrix $B_{jk}$ is calculated by matching the Maclaurin series for the BGL prefactors and the BCL pole \cite{BCL}. The matrix for $B_{jk}$ in $B\rightarrow D \ell \nu$ is provided in Tab. I.
Following~\cite{BCL}, we choose the outer function to be $\phi_+(z) = 1$, and the Blaschke factor to be $P_+(z) = 1 - q^2(z)/m_{B_c^*}^2$. The free parameters of this parameterization are the $b_{+,n}$ terms, and there are $N$ of them. The BCL parameterization uses a $z$-expansion with $q^2$ instead of $w$, which has the form 

\begin{equation} 
    \begin{split}
        z(q^2) &= \frac{\sqrt{t_+ - q^2} - \sqrt{t_+ - t_0}}{\sqrt{t_+ - q^2} + \sqrt{t_+ - t_0}}
    \end{split}    
\end{equation}
Where $t_+$ and $t_0$ are constants defined by $t_+ = (m_B+m_D)^2$ and $t_0 = (m_B+m_D) (\sqrt{m_B} - \sqrt{m_D})^2$. The choice of $t_0$ ensures that that this expansion converges. 

\begin{table}[b]
    \centering
    \begin{tabular}{ccccccc}
    \hline
         $B_{00}$ & $B_{01}$
         & $B_{02}$
         & $B_{03}$
         & $B_{04}$& $B_{05}$\\\hline
        0.0197 & 0.0042 & -0.0109 & -0.0059 & -0.0002 & 0.0012\\
        \hline\hline
    \end{tabular}
    \caption{Calculated elements of the $B_{jk}$ matrix}
    \label{tab:my_label}
\end{table}

The final parameterization considered was the CLN parameterization. The advantage of this parameterization is that it reduces the number of free parameters, but it does this by adding dispersive constraints and symmetries which make this parameterization a model-dependent one. The CLN parameterization has a form factor that looks like,

\begin{equation}
\begin{split}
    G(z) = G(1) (&1 - \rho^2 z + (51 \rho^2 - 10) z^2 - (252 \rho^2 - 84) z^3)
\end{split}
\end{equation}
\noindent
The free parameters are $G(1)$ and $\rho^2$.

\section{Results}
\label{sec:results}
\noindent
We compare these different parameterizations of the form factor by fitting the free parameters to differential decay data obtained by the Belle Collaboration~\cite{mainb2d}. We perform these fits using a limited number of data points in order to test how well these parameterizations are able to predict the remaining data points. We check the accuracy of the fits by finding the reduced-$\chi^2$ values. We compare the fitted and predicted reduced-$\chi^2$ values of the parameterizations.

The reduced-$\chi^2$ values that we calculated are of the standard form,
\begin{equation}
    \chi^2_{reduced} = \frac{1}{N} \sum_{i} \frac{(O_i - C_i)^2}{\sigma^2_i}
\end{equation}
\noindent
Where $O_i$ is the data collected by the Belle collaboration, $C_i$ is the results from our fits, and $\sigma^2_i$ are the uncertainties in $O_i$. We considered two different $\chi^2_{reduced}$, one for the region that was included in the fit $\chi^2_{reduced,fitted}$ and one for the region that was not included in the fit $\chi^2_{reduced,predicted}$. In the fitted region, $N$ is the number of degrees of freedom $N_{dof}$ and the sum over $i$ is over the data points used in the fit. Whereas, for the predicted region, $N$ is the number of data points not used in the fit and the sum over $i$ is over the data points not used in the fit. The fits were performed using the $lsqfit$ python library, which implements non-linear least square fitting methods to minimize the $\chi^2$ function~\cite{lsqfit}.

\begin{figure}
\begin{center}
        \includegraphics[scale=0.33]{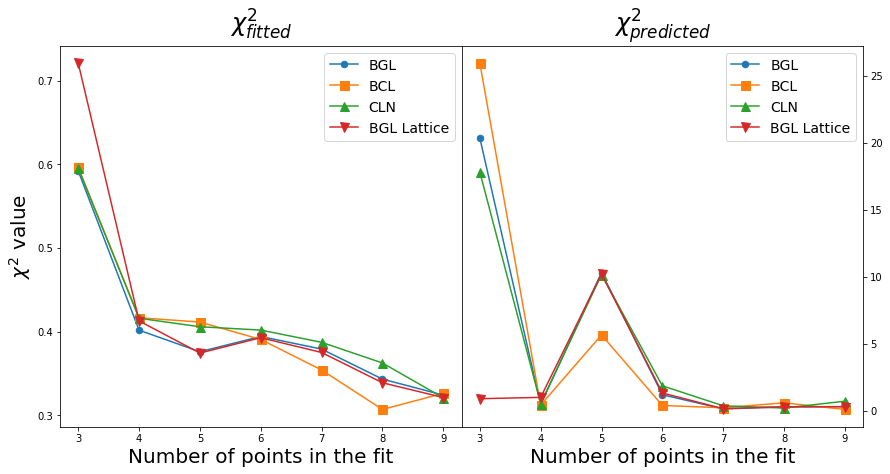}
\end{center}
        \caption{The $\chi^2_{reduced}$ values for each fit of all parameterizations used in this analysis}
        \label{figure:1}
\end{figure}

We start by fitting the first $n$ data points and predict the remaining $10-n$ data points, where $3 \leq n \leq 9$. Comparing our fits to the data points included in the fit produce $\chi^2_{reduced,fitted}$, and comparing our fits with the data points not included in the fits produce $\chi^2_{reduced,predicted}$. The results of all $\chi^2_{reduced}$ can be seen in \ref{figure:1}.

The red triangles in Fig. \ref{figure:1} used the results from lattice calculations~\cite{Bailey_2015pi} as the priors. Bailey et. al. used lattice calculations to find the free parameters of the BGL parameterization, which are the results we use. The lattice priors over-constraining our fit, so their uncertainties are scaled upwards by a factor of 30 to allow the fit more freedom. 

\subsection{BGL Parameterization}

\begin{figure}[H]
\begin{center}
        \includegraphics[scale=0.5]{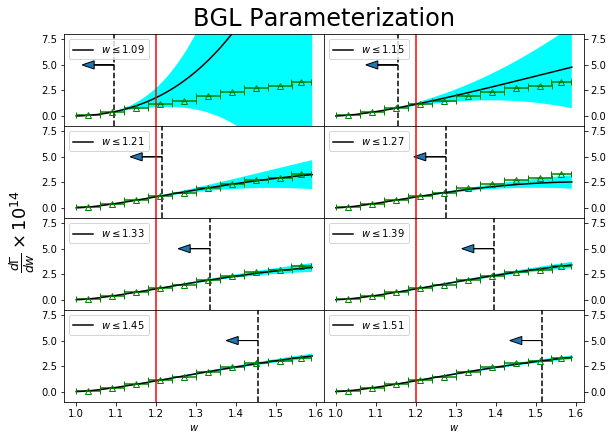}
\end{center}
        \caption{Results of fitting the BGL parameterization to different numbers of data points}
        \label{figure:2}
\end{figure}

\noindent
 The results for the BGL parameterization from Fig. \ref{figure:1} can be seen graphically in Fig. \ref{figure:2}. In these plots the green triangles are the Belle collaboration's data with error bars, the black line is the function with the fitted parameters as inputs, the teal region is the 1-$\sigma$ error band, and the vertical dashed line with arrow pointing to the left indicates the highest $w$-value used in the fit. The red line corresponds to the lattice regime, the largest $w$-value where lattice results are obtainable, $w < 1.2$~\cite{Bailey_2015pi}.

We found that the results for the BGL parameterization with more than two parameters have significantly larger $\chi^2_{reduced,predicted}$ for all regions, so we only consider this parameterization with two free parameters.

\subsection{BCL Parameterization}

\begin{figure}[H]
\begin{center}
        \includegraphics[scale=0.5]{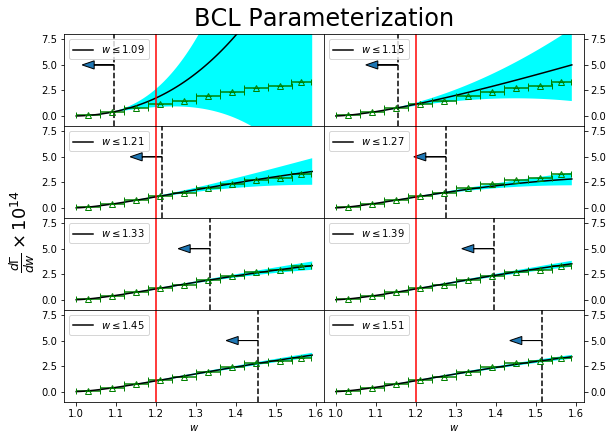}
\end{center}
        \caption{The same as Fig. \ref{figure:2} but for the BCL parameterization}
        \label{figure:3}
\end{figure}

\noindent
We performed the same procedure for the BCL parameterization that we did for the BGL parameterization, with the results being shown graphically in Fig. \ref{figure:3}. 

We found that the $\chi^2_{reduced}$ values for the BCL parameterization with three parameters was identical to using only two parameters, and we found that for more than three parameters the $\chi^2_{reduced,predicted}$ was significantly larger for all regions, so we only considered the BCL parameterization with two free parameters.

\subsection{CLN Parameterization}

\begin{figure}[H]
\begin{center}
        \includegraphics[scale=0.5]{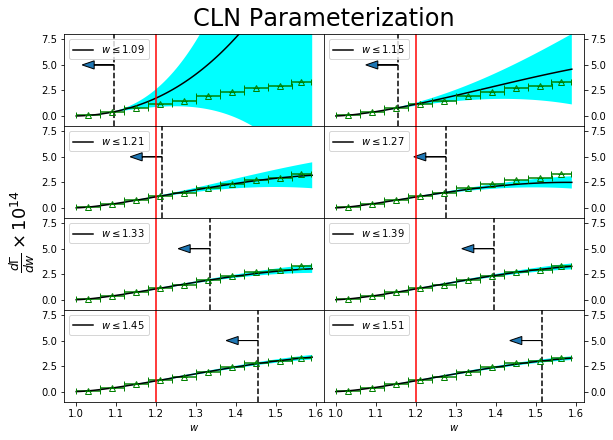}
\end{center}
        \caption{The same as Fig. \ref{figure:2} but for the CLN parameterization}
        \label{figure:4}
\end{figure}

\noindent
The final parameterization that was considered was the CLN parameterization, which has a fixed number of free parameters. The model dependence of the CLN parameterization makes it less ideal than the BGL and BCL parameterizations which are both model independent. The results for this parameterization can be seen in Fig. \ref{figure:4}.

\subsection{Using Lattice Data as a Prior}

\begin{figure}[H]
\begin{center}
        \includegraphics[scale=0.5]{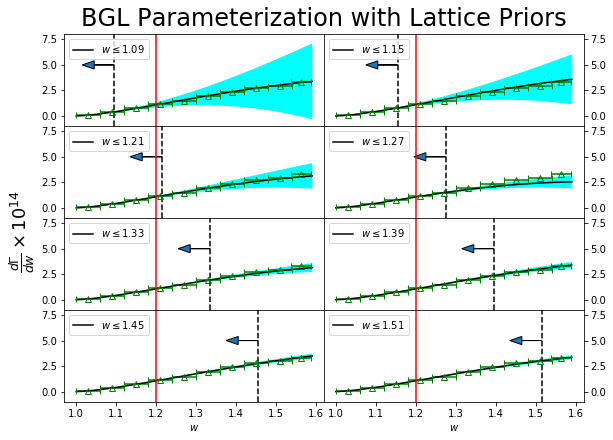}
\end{center}
        \caption{The same as \ref{figure:2} but for the BGL parameterization with lattice priors}
        \label{figure:5}
\end{figure}

\noindent
The final variation that we considered was to use the results from lattice calculations~\cite{Bailey_2015pi} as the priors. Bailey et. al. used lattice calculations to find the free parameters of the CLN and BGL parameterizations, although they only report their results for BGL. Again, only considering the BGL parameterization for two parameters, our results can be seen graphically in Fig. \ref{figure:5}. 

\section{Conclusion}
\noindent
We investigated three different parameterizations of the vector form factor describing the decay rate of the process $B \rightarrow D l \nu_l$. Our goal was to determine which of these parameterizations would most accurately predict the high energy regime from fits performed in the low energy regime, and we found the $\chi^2$-values of these fits in order to determine their accuracy. Performing these fits in the low $w$ regime is a useful replica of the conditions of lattice gauge theorists for us to determine the best form factor for them to use.

By comparing the different $\chi^2_{reduced,predicted}$ for the different parameterizations at low $w$, it appears that the BCL parameterization provides slightly better predictions. We also looked at how inputting lattice priors would affect our fits, and even with significantly increased uncertainties these lattice priors fit and predict the data with exceptional accuracy. 

\subsection*{Acknowledgments}
\noindent
This work was supported by DOE grant No. DE-SC0010113.

\bibliographystyle{JHEP}
\bibliography{proceedings}

\end{document}